\documentclass[aps,prl,twocolumn,superscriptaddress]{revtex4-1}

\usepackage{booktabs}
\usepackage{array}
\usepackage{graphicx}
\usepackage{dcolumn}
\usepackage{bm}
\usepackage{amsmath}
\usepackage{color}
\usepackage{verbatim}
\usepackage{natbib} 
\usepackage{verbatim}
\usepackage[colorlinks,
            linkcolor=blue,
            anchorcolor=blue,
            citecolor=blue,
urlcolor=blue
            ]{hyperref}
\usepackage{color}
\usepackage{comment}
\usepackage{ulem}

\newcommand{\RNum}[1]{\uppercase\expandafter{\romannumeral#1\relax}}

\usepackage{bm}

\begin{document}


\title{Direct Characterization of Quantum Measurements using Weak Values}


\author{Liang Xu}
\affiliation{National Laboratory of Solid State Microstructures, Key Laboratory of Intelligent Optical Sensing and Manipulation, College of Engineering and Applied Sciences, and Collaborative Innovation Center of Advanced Microstructures, Nanjing University, Nanjing 210093, China}
\affiliation{Research Center for Quantum Sensing, Zhejiang Lab, Hangzhou 310000, China}
\author{Huichao Xu}
\affiliation{National Laboratory of Solid State Microstructures, Key Laboratory of Intelligent Optical Sensing and Manipulation, College of Engineering and Applied Sciences, and Collaborative Innovation Center of Advanced Microstructures, Nanjing University, Nanjing 210093, China}
\affiliation{Purple Mountain Laboratories, Nanjing, Jiangsu 211111, China}
\author{Tao Jiang}
\affiliation{National Laboratory of Solid State Microstructures, Key Laboratory of Intelligent Optical Sensing and Manipulation, College of Engineering and Applied Sciences, and Collaborative Innovation Center of Advanced Microstructures, Nanjing University, Nanjing 210093, China}
\author{Feixiang Xu}
\affiliation{National Laboratory of Solid State Microstructures, Key Laboratory of Intelligent Optical Sensing and Manipulation, College of Engineering and Applied Sciences, and Collaborative Innovation Center of Advanced Microstructures, Nanjing University, Nanjing 210093, China}
\author{Kaimin Zheng}
\affiliation{National Laboratory of Solid State Microstructures, Key Laboratory of Intelligent Optical Sensing and Manipulation, College of Engineering and Applied Sciences, and Collaborative Innovation Center of Advanced Microstructures, Nanjing University, Nanjing 210093, China}
\author{Ben Wang}
\affiliation{National Laboratory of Solid State Microstructures, Key Laboratory of Intelligent Optical Sensing and Manipulation, College of Engineering and Applied Sciences, and Collaborative Innovation Center of Advanced Microstructures, Nanjing University, Nanjing 210093, China}
\author{Aonan Zhang}
\affiliation{National Laboratory of Solid State Microstructures, Key Laboratory of Intelligent Optical Sensing and Manipulation, College of Engineering and Applied Sciences, and Collaborative Innovation Center of Advanced Microstructures, Nanjing University, Nanjing 210093, China}
\author{Lijian Zhang}
\email{lijian.zhang@nju.edu.cn}
\affiliation{National Laboratory of Solid State Microstructures, Key Laboratory of Intelligent Optical Sensing and Manipulation, College of Engineering and Applied Sciences, and Collaborative Innovation Center of Advanced Microstructures, Nanjing University, Nanjing 210093, China}


\date{\today}

\begin{abstract}
{The time-symmetric formalism endows the weak measurement and its outcome, the weak value, many unique features. In particular, it allows a direct tomography of quantum states without resort to complicated reconstruction algorithms and provides an operational meaning to wave functions and density matrices. Here, we propose and experimentally demonstrate the direct tomography of a measurement apparatus by taking the backward direction of weak measurement formalism. Our protocol works rigorously with the arbitrary measurement strength, which offers an improved accuracy and precision. The precision can be further improved by taking into account the completeness condition of the measurement operators, which also ensures the feasibility of our protocol for the characterization of the arbitrary quantum measurement. Our work provides new insight on the symmetry between quantum states and measurements, as well as an efficient method to characterize a measurement apparatus.}
\end{abstract}


\maketitle
\textit{Introduction.}-- A time-symmetric description of a physical system involving both the initial and final boundary conditions allows not only the prediction but also retrodiction of its evolution, therefore may reveal more information about the system. Such description for quantum systems can be captured by the two-state vector formalism (TSVF) \cite{aharonov2008two, aharonov1991complete}. The pre-selected state describing the preparation of the system evolves forward in time, while the post-selected state determined by the measurement on the system evolves backward. The TSVF provides a new approach to interpret many intriguing quantum effects \cite{hardy1992quantum, resch2004experimental, lundeen2009experimental, aharonov2013quantum, Yokota_2009, kim2020observing, palacios2010experimental}. In particular, adding retrodiction to the standard predictive approach allows an improved description of the evolution trajectories of quantum systems \cite{PhysRevLett.111.240402, PhysRevLett.111.160401, weber2014mapping, PhysRevLett.114.090403}.

One of the most remarkable phenomena from TSVF is the weak measurement and its associated outcome, the weak value \cite{aharonov1988result}, which are widely used in precision metrology \cite{hosten2008observation, dixon2009ultrasensitive, strubi2013measuring, magana2014amplification, xu2013phase} and investigations of various phenomena \cite{sjoqvist2006geometric, kobayashi2010direct, kobayashi2011observation, cho2019emergence, huang2019simulating}. In particular, the complex weak value can be the complex probability amplitude of the wavefunction, therefore allows a direct tomography of quantum states and processes \cite{lundeen2012procedure, salvail2013full, thekkadath2016direct, wu2013state, malik2014direct, pan2019direct, vallone2016strong, zhu2016direct, zhang2018direct, zhu2019hybrid, yang2019zonal, kim2018direct, shikano2009weak, kobayashi2014stereographical, PhysRevLett.127.040402}. Compared to the conventional tomography scheme that reconstructs a quantum state with an overcomplete set of measurements followed by the complex post-processing of data \cite{banaszek1999maximum, lvovsky2009continuous, zhang2012mapping}, the direct tomography avoids the reconstruction algorithm and shows distinct advantages in both directness and simplicity. Therefore, the direct tomography promises to be especially useful in the characterization of high-dimensional states \cite{malik2014direct, PhysRevLett.127.040402}.

Until now direct tomography only takes the forward direction of the weak measurement to characterize the pre-selected state and its evolution. The time-symmetric formulation has not been fully explored. Since the pre- and post-selected states enter the formulation on equal footing, it is expected that the backward direction allows to directly determine the post-selected state, which can be viewed as the retrodicted state of the measurement performed on the quantum system \cite{amri2011characterizing}. This connection implies the feasibility of direct tomography of a quantum measurement \cite{PhysRevA.98.042318}. 

In this paper, we propose the general framework for the direct quantum detector tomography (DQDT) protocol. A direct connection between the numerator of the weak value and either the eigenvector of the rank-1 positive operator-valued measure (POVM) or the Dirac distribution of the higher-rank POVM is rigorously established with the arbitrary measurement strength. Such connection allows to not only directly determine the POVM element of interest but also improve the accuracy and precision by adjusting the measurement strength. As a demonstration, we experimentally characterize both projective measurements and general POVMs in the polarization degree of freedom (DOF) of photons. Moreover, we show the feasibility of our protocol for the characterization of the arbitrary POVM thanks to the completeness condition of the POVM.

\begin{figure}[ht]
\centering
\includegraphics[width=0.47\textwidth]{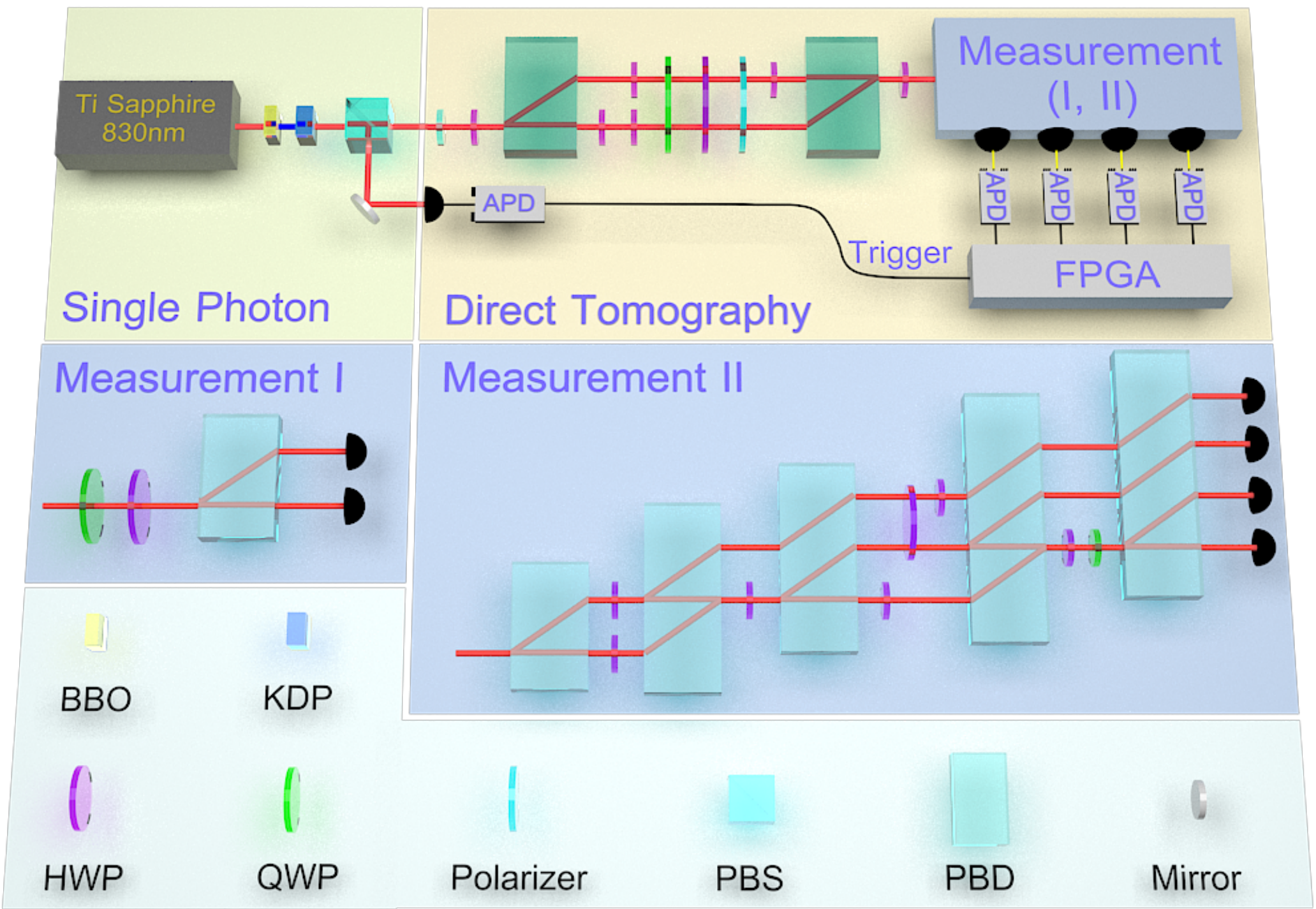}
\caption{Experimental Setup. (a) The pulsed laser at 830nm first gets through a $\beta$-barium borate (BBO) crystal for the second harmonic generation and then inputs the potassium di-hydrogen phosphate (KDP) crystal for spontaneous parametric downconversion (SPDC). Two photons are simultaneously generated in SPDC, one of which is used for heralding and the other is collected to input the 'Direct Tomography' module for the characterization of the unknown quantum measurement labelled `Measurement (\RNum{1}, \RNum{2})'. The modules `Measurement \RNum{1}' and `Measurement \RNum{2}' illustrate the experimental implementation of projective measurements and symmetric informationally complete positive-operator-valued measure (SIC POVM) in the polarization degree of freedom, respectively.
}\label{Expsetup}
\end{figure}

\textit{Theoretical framework of DQDT.}--The original direct tomography protocol is applied to determining a pure quantum state $|s\rangle$ which is expanded in an orthogonal basis $\{|j\rangle\} (\mathcal{J})$ with the probability amplitudes $\{\alpha_j\}$. To directly obtain $\alpha_j = \langle j|s\rangle$, the unknown state $|s\rangle$ is weakly measured with the projector $\hat{\pi}_j = |j\rangle\langle j|$ of the orthogonal base state $|j\rangle$ followed by the post-selection to $|\psi_n\rangle \propto \sum_j |j\rangle$. The outcome of this measurement is the weak value of $\hat{\pi}_j$, given by $\langle \hat{\pi}_j^w\rangle_{s}^{(n)} = \langle \psi_n|\hat{\pi}_j|s\rangle/\langle \psi_n|s\rangle\propto \alpha_j$. Thus, by measuring $\langle \hat{\pi}_j^w\rangle_{s}^{(n)}$ that scans in the basis $\mathcal{J}$, $\{\alpha_j\}$ can be completely determined with the normalization of the state. The direct tomography protocol is also extended to mixed states by exploiting the connection between the weak value and the Dirac distribution or the density matrix \cite{lundeen2012procedure, salvail2013full, thekkadath2016direct}.

As the formulation of weak value is symmetric for both pre- and post-selected states ($i.e.$, $|s\rangle$ and $|\psi_n\rangle$), the post-selected state $|\psi_n\rangle$ can also be directly measured by properly preparing the pre-selected state $|s\rangle$. The post-selection to $|\psi_n\rangle$ is typically implemented with a projective measurement $\hat{\Pi}_n = |\psi_n\rangle\langle \psi_n|$. Therefore, this direct method is expected to be able to characterize a quantum measurement. 

We first concentrate on characterizing one element of a rank-1 POVM $\hat{\Pi}_n = \eta_n |\psi_n\rangle\langle \psi_n|$, in which $|\psi_n\rangle$ can be considered as the retrodicted state and $\eta_n$ is the equivalent detection efficiency \cite{barnett2000bayes, amri2011characterizing, DQDT_supplementary, jabbour2018multiphoton}. The pre-selected state of the quantum system (QS) is prepared to $\rho_s = |s\rangle\langle s|$ with $|s\rangle \propto \sum_j |j\rangle$. We implement the Von Neumann measurement of $\rho_s$ by coupling the QS to a meter state (MS) $\rho_m$ under the Hamiltonian $\hat{H} = g\delta(t-t_0)\hat{\pi}_j \hat{M}$ giving the joint state $\rho_{jt} = \exp[-i\int \hat{H} dt ]\rho_s\otimes \rho_m \exp[i\int \hat{H} dt]$, where $g$ is the coupling strength and $\hat{M}$ is the observable of the MS. Finally, the unknown measurement operator $\hat{\Pi}_n$ performs the post-selection on the QS. The generalized weak value formulated as
\begin{equation}
\langle \hat{\pi}_j^w\rangle_s^{(n)} = \frac{\text{Tr}(\hat{\Pi}_n \hat{\pi}_j\rho_s)}{\text{Tr}(\hat{\Pi}_n \rho_s)}
\end{equation}
can be extracted from the measurement of the MS that survives the post-selection. To simplify the process in determining $\hat{\Pi}_n$, we define an unnormalized state $|\varphi_n\rangle = \sqrt{\eta_n}|\psi_n\rangle$, thus $\hat{\Pi}_n = |\varphi_n\rangle\langle \varphi_n|$. Instead of successively determining $\eta_n$ and $|\psi_n\rangle$, we directly measure $|\varphi_n\rangle$ to obtain $\hat{\Pi}_n$. Since there is no normalization condition in $|\varphi_n\rangle$, measuring the numerator of weak value, $i.e.$, $\omega_{j,s}^{(n)} = \text{Tr}(\hat{\Pi}_n\hat{\pi}_j\rho_s)$, is sufficient for characterizing $\hat{\Pi}_n$. In this way, we have $\omega_{j,s}^{(n)} = \langle s|\varphi_n\rangle\langle \varphi_n|j\rangle\langle j|s\rangle = t_{s}^{(n)}\langle \varphi_n|j\rangle/\sqrt{d}$, where $t_s^{(n)}=\langle s|\varphi_n\rangle$ and $d = 1/|\langle j|s\rangle|^2$ is the dimension of the QS. Since $t_s^{(n)}$ is a constant for different $j$, we take $t_s^{(n)}$ as a real value that can be derived by $t_s^{(n)} = \sqrt{\langle s|\varphi_n\rangle\langle \varphi_n|s\rangle} = \sqrt{\text{Re}(\sum_j \omega_{j,s}^{(n)})}$ (taking the real part for practical data processing). Consequently, each expanding coefficient of $|\varphi_n\rangle$ in the base state $|j\rangle$ is given by
\begin{equation}
\langle j|\varphi_n\rangle = \frac{\sqrt{d}\omega_{j,s}^{(n)*}}{\sqrt{\text{Re}(\sum_j \omega_{j,s}^{(n)})}}.
\label{Eq:rank-1Tomo}
\end{equation}
In the general situation to directly characterize higher-rank POVM element $\hat{\Pi}_n$, $\omega_{j,s}^{(n)}$ naturally equals to the `right' phase-space Dirac distribution of $\hat{\Pi}_n$ in the $d$-dimensional discrete Hilbert space by scanning both the projectors $\hat{\pi}_j$ in the basis $\mathcal{J}$ and the pre-selected state $|s\rangle = 1/\sqrt{d}\sum_{j=0}^{d-1} e^{2\pi i j s/d}|j\rangle$ in the Fourier basis $\mathcal{S}$ \cite{chaturvedi2006wigner}.

For a qubit MS initialized as $\rho_m = |0\rangle\langle 0|$ and the observable $\hat{M} = \hat{\sigma}_y$, both the real and imaginary parts of $\omega_{j,s}^{(n)}$ can be obtained from the joint measurement of the QS with the post-selection operator $\hat{\Pi}_n$ and the MS with the observables $\hat{S}_1 = \hat{\sigma}_x + \tan(g/2)(\hat{\openone} - \hat{\sigma}_z)$ and $\hat{S}_2 = \hat{\sigma}_y$, formulated as \cite{DQDT_supplementary}
\begin{equation}
\omega_{j,s}^{(n)} = \frac{1}{2\sin(g)}\text{Tr}[\rho_{jt}\hat{\Pi}_n\otimes (\hat{S}_1 + i\hat{S}_2)].
\label{Eq:S_observable}
\end{equation}

\begin{figure*}[ht]
\centering
\includegraphics[width=0.95\textwidth]{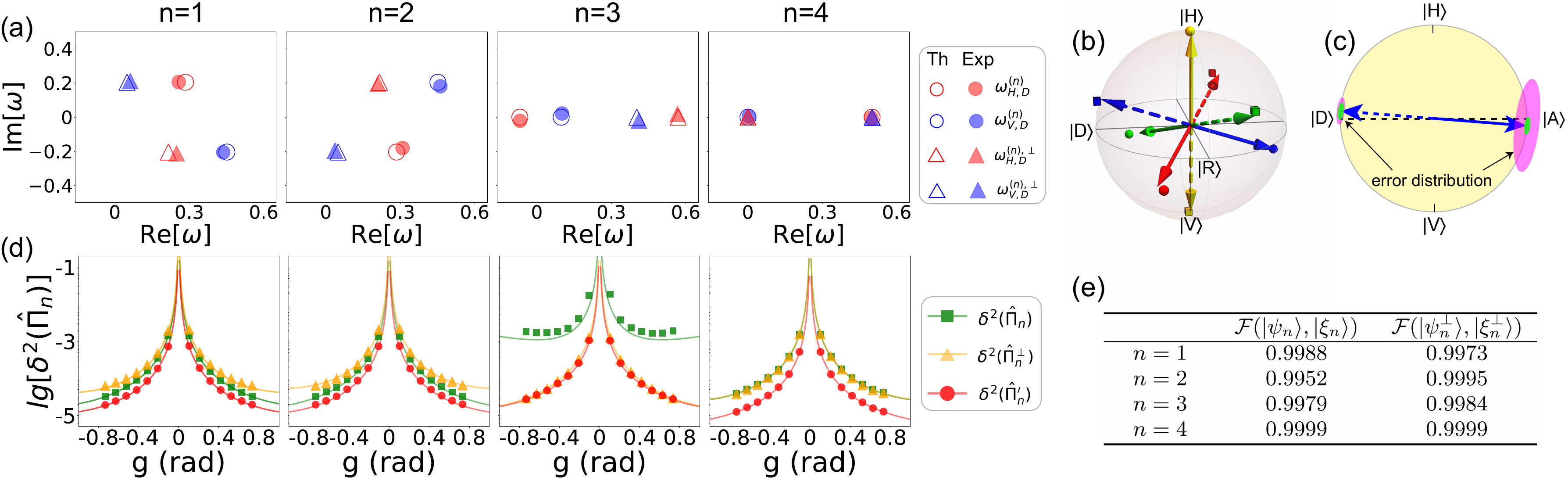}
\caption{The experimental results for the characterization of the `Measurement \RNum{1}'. (a) We plot the experimentally (points) estimated and theoretical (circles) $\omega_{j,s}^{(n)}$ and $\omega_{j,s}^{(n),\perp}$ by denoting the real and the imaginary parts of $\omega$ as the horizontal and the vertical axises, respectively. (b) In the Bloch sphere representation, the solid (dashed) arrows refer to the theoretical retrodicted states $|\xi_n\rangle$ ($|\xi_n^\perp\rangle$) and the points (cubes) represent the experimental retrodicted states $|\psi_n\rangle$ ($|\psi_n^\perp\rangle$). The red, green, blue and yellow colors refer to $n=1,2,3,4$, respectively. (c) The schematic error distributions of the measured $|\psi_3\rangle$ and $|\psi_3^\perp\rangle$ before (pink shaded area) and after (green shaded area) the normalization are shown in the cross-section of Bloch sphere. (d) The precision of the experimentally measured POVM (points) obtained from the Monte Carlo simulation with different $g$ are compared with the theoretical results (lines). The $\delta^2(\hat{\Pi}_n)$ and $\delta^2(\hat{\Pi}^\perp_n)$ denote the variances of the POVM elements before the normalization. The variance $\delta^2(\hat{\Pi}^\prime_n)$ for both the POVM elements ($\hat{\Pi}_n$ and $\hat{\Pi}^\perp_n$) after the normalization are the same. (e) The fidelities ($\mathcal{F} = |\langle\psi|\phi\rangle|^2$ for $|\psi\rangle$ and $|\phi\rangle$) between the experimental retrodicted states $|\psi_n\rangle$ ($|\psi_n^\perp\rangle$) and the theoretical states $|\xi_n\rangle$ ($|\xi_n^\perp\rangle$) for $n=1,2,3,4$ are given in the table.
}\label{proj_rank1}
\end{figure*}

The feasibility of a tomography technique is typically evaluated by the metrics of both accuracy and precision. In the original direct state tomography, the accurate weak value can only be obtained with a small coupling strength due to the first-order approximation, which sacrifices precision. In view of effective efforts made in the state tomography  \cite{vallone2016strong, denkmayr2017experimental, calderaro2018direct}, the DQDT protocol is valid for the arbitrary non-zero $g$ allowing us to improve the precision of the tomography without loss of accuracy. Besides, the potential inefficacy of direct state tomography is caused by the orthogonality of the pre- and post-selected states \cite{PhysRevA.84.052107, PhysRevA.89.022122, PhysRevA.92.062133}. In the following, we show this efficacy issue can be overcome in the DQDT by the appropriate use of the completeness condition of the POVM.

The precision of the measured POVM $\{\hat{\Pi}_n\}$ is quantified by summing the variance $\delta^2 (\cdot)$ of each POVM element $\Delta^2 = \sum_n \delta^2(\hat{\Pi}_n)$, where $\delta^2 (\hat{\Pi}_n) = \sum_{j,j^\prime} \delta^2 (\langle j|\hat{\Pi}_n|j^\prime\rangle)$ with $|j\rangle$ and $|j^\prime\rangle$ the base states in basis $\mathcal{J}$. Since each POVM element $\hat{\Pi}_n$ is independently determined in the DQDT, the precision $\delta^2 (\hat{\Pi}_n)$ can be improved when the completeness condition of POVM $i.e.$, $\sum_n \hat{\Pi}_n = \hat{\openone}$ is applied. To acquire the optimal precision based on the completeness condition, the normalized matrix entries of the POVM element $\hat{\Pi}_n$ can be obtained by a weighted average of the directly measured results and those inferred from other POVM elements. The optimal weighting factors should be proportional to their metrological contributions quantified by the inverse of their variances \cite{DQDT_supplementary}. As an example, we let $E_k = \text{Re}(\langle j|\hat{\Pi}_k|j^\prime\rangle)$ and $I_k = 1/\delta^2(E_k)$. The normalized $E^\prime_k$ according to the completeness of POVM is given by
\begin{equation}
E^\prime_k = \frac{I_k}{I_k+I^\circ_k}E_k + \frac{I^\circ_k}{I_k+I^\circ_k}(\delta_{j,j^\prime}-\sum_{m\neq k}^M E_m),
\label{Eq:weighted_ave}
\end{equation}
where $I^\circ_k = 1/[\sum_{m\neq k}^M \delta^2 (E_m)]$ and $\delta_{j,j^\prime}$ denotes the Kronecker delta function. For the general POVM element, the positive condition ($\hat{\Pi}_n \ge 0$) is also useful to improve the precision \cite{DQDT_supplementary}.

\textit{Experiment.}-- The experimental setup is shown in Fig. \ref{Expsetup}. By referring to the polarization DOF of photons as the QS, we prepare the pre-selected state through a polarizer and a half-wave plate (HWP). Then, the pre-selected photons input a polarizing beam displacer (PBD), which converts the polarization-encoded QS to path-encoded. The polarization of photons in the two paths is employed as the MS. We use the computational basis $\{|0\rangle,|1\rangle\}$ referring to the $|H\rangle$ and $|V\rangle$ for polarization qubit or the upper and lower path for the path qubit. A HWP at $45^\circ$ placed in path `1' initializes the MS to $|0\rangle_m$. The coupling Hamiltonian $\hat{H} = g\delta(t-t_0)\hat{\pi}_j\hat{\sigma}_y$ is realized by rotating $g/2$ degrees the HWP placed in $j$th path. Since the post-selection on the QS and the measurement of the MS are physically commutable, we first perform the projective measurement on the MS with the following combination of a quarter-wave plate (QWP), a HWP and a polarizer \cite{Chen:s}. Afterwards, the transmitted photons are recombined with the subsequent HWP at $45^\circ$ compensating the changes of the polarization during the coupling process. Finally, the quantum measurement, which is to be characterized, implements the post-selection on the QS.

\begin{figure*}[ht]
\centering
\includegraphics[width=0.9\textwidth]{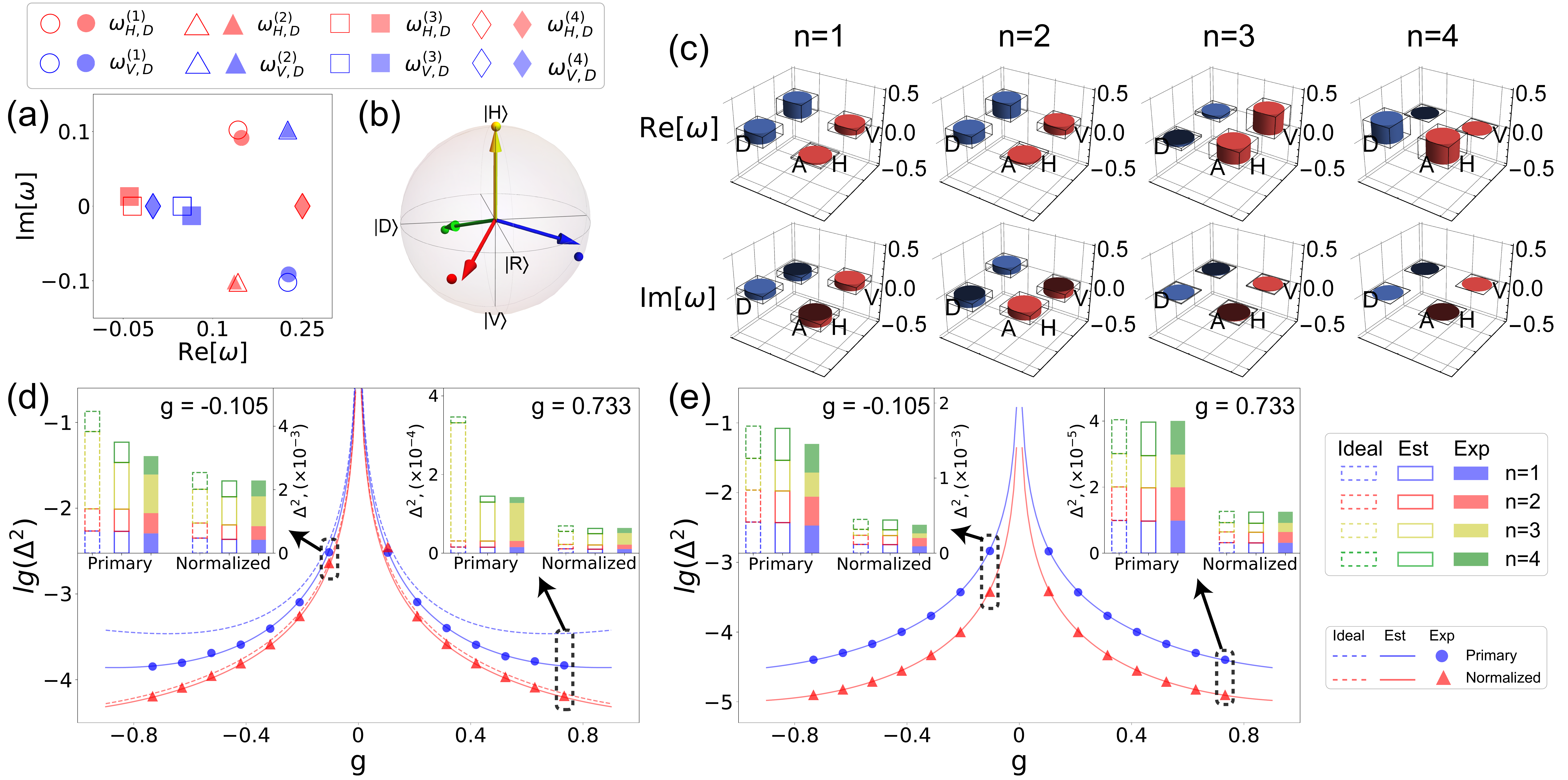}
\caption{The experimental results for the characterization of the `Measurement \RNum{2}'. (a) By regarding the `Measurement \RNum{2}' as rank-1 POVM, we illustrate the experimentally (solid markers) estimated and theoretical (hollow markers) $\omega_{j,s}^{(n)}$. (b) The experimentally derived pure retrodicted states (points) are compared with those of the ideal SIC POVM (arrows) in the Bloch sphere with fidelities 0.9781 (red), 0.9912 (green), 0.9882 (blue), 0.9984 (yellow) for $n=1,2,3,4$, respectively. (c) The experimentally estimated (the cylinders) Dirac distributions of the `Measurement \RNum{2}' are compared with those inferred from the results of conventional tomography (CT) (solid edges). The fidelities ($\mathcal{F} = [\text{Tr}(\sqrt{\sqrt{\rho_1}\rho_2\sqrt{\rho_1}})]^2$ for $\rho_1$ and $\rho_2$) between the retrodicted states of the directly measured POVM with those of CT are 0.9994, 0.9994, 0.9999, 0.9997 for $n=1,2,3,4$, respectively. The precision of the obtained POVM is illustrated in (d) (rank-1 situation) and (e) (rank-2 situation). In (d), we compare the precision of the experimentally measured POVM (points and triangles) obtained form the Monte Carlo simulation with the calculated precision of the ideal SIC POVM (dashed lines) and the estimated POVM (lines). In (e), the calculated precision inferred from the results of the CT (lines) and the ideal SIC POVM (dashed lines) coincide with each other. In both (d) and (e), $\Delta^2$ in $\text{lg}(\Delta^2)$ of the $y$-axis, represents the total variance of the POVM. The two insets show the precision of each POVM elements for $g=-0.105$ and $g = 0.733$, respectively. Here, we denote the precision of the directly measured results as `Primary' (blue) and that of the normalized results as `Normalized' (red).
}\label{sic_tomo_precision}
\end{figure*}

The `Measurement \RNum{1}', composed of a QWP, a HWP and a PBD, performs the projective measurement on the polarization of photons. We implement four configurations of projective measurements $\{|\xi_n\rangle\langle \xi_n|,|\xi_n^\perp\rangle\langle \xi_n^\perp|\}(n=1,2,3,4)$, in which the retrodicted states are parameterized as $|\xi_n\rangle = a_n|H\rangle + b_n e^{i\phi_n}|V\rangle$ and $|\xi_n^\perp\rangle = b_n|H\rangle - a_n e^{i\phi_n}|V\rangle$ with the parameters $2a_1 = 2a_2 = 2a_3 = b_1 = b_2 = b_3 = 2/\sqrt{3}$, $\phi_1 = \pi/3$, $\phi_2 = -\pi/3$, $\phi_3 = \phi_4 = 0$, $a_4 = 1$, $b_4 = 0$. Since the retrodicted states are pure, we fix the pre-selected state to $|s\rangle = |D\rangle = 1/\sqrt{2}(|H\rangle + |V\rangle)$. The coupling strength $g$ is varied from $-42^\circ$ to $42^\circ$ with the step of $6^\circ$. 

Based on the Eq. \eqref{Eq:S_observable}, the measurement results of the observable $\hat{S}_1$ and $\hat{S}_2$ for different $g$ are fitted to obtain an estimated $\omega_{j,s}^{(n)}$ shown in Fig. \ref{proj_rank1} (a) \cite{DQDT_supplementary}. The corresponding retrodicted states of the four projective measurements are compared with the theoretical results in Fig. \ref{proj_rank1} (b). The fidelities between the experimental ($|\psi_n\rangle$ and $|\psi_n^\perp\rangle$) and the theoretical ($|\xi_n\rangle$ and $|\xi_n^\perp\rangle$) retrodicted states are given in Table (e) of Fig \ref{proj_rank1}. Fig. \ref{proj_rank1} (c) schematically illustrates the error distribution of the projective measurements $|\xi_3\rangle\langle \xi_3|$ and $|\xi_3^\perp\rangle\langle \xi_3^\perp|$ both before and after the normalization of the POVM. Since $|\langle D|\xi_3\rangle|^2$ is close to 0, the precision of determining $|\xi_3\rangle$ is originally much worse than that of $|\xi_3^{\perp}\rangle$, which is significantly improved after the normalization. The detailed results of precision for different $g$ are shown in Fig. \ref{proj_rank1} (d). As we can see, the weighted average in Eq. \eqref{Eq:weighted_ave} ensures that the precision of the normalized projector is better than that of either one in the arbitrary two-output projective measurement.

The `Measurement \RNum{2}' in Fig. \ref{Expsetup} illustrates the realization of the symmetric informationally complete (SIC) POVM on the polarization DOF of photons through quantum walk \cite{bian2015realization, zhao2015experimental}. Ideally, the POVM elements of the four outputs from the bottom to the top are respectively $\hat{\Pi}_n = \eta_n|\xi_n\rangle\langle \xi_n|$ for $n$ from 1 to 4 with all the equivalent detection efficiency $\eta_n = 0.5$. Since all the retrodicted states are pure, we first adopt the same DQDT procedure as the projective measurement. The experimental $\omega_{j,s}^{(n)}$ estimated from the results of different $g$ and the derived retrodicted states are shown in Fig. \ref{sic_tomo_precision} (a) and (b), respectively. The precision of the SIC POVM both before and after the normalization for different $g$ is illustrated in Fig. \ref{sic_tomo_precision} (d). 

Due to the imperfect optical interference caused by the spatial misalignment and air turbulence, the retrodicted states of the realistic $\hat{\Pi}_n$ may be mixed states. Therefore, the above results acquired by regarding the `Measurement \RNum{2}' as rank-1 POVM may deviate from the actual situation. To remove this error, we also prepare the pre-selected state $|A\rangle = 1/\sqrt{2}(|H\rangle - |V\rangle)$ to acquire the full Dirac distribution of all the POVM elements $\omega_{j,s}^{(n)}$ ($j = H,V$ and $s = D,A$) with different $g$, respectively. As a comparison, we reconstruct the POVM of the `Measurement \RNum{2}' by the conventional tomography (CT) \cite{DQDT_supplementary}. The directly measured Dirac distribution that are estimated from the results of different $g$ are compared with those derived from the results of CT, shown in Fig. \ref{sic_tomo_precision} (c). The precision of the DQDT in characterizing the `Measurement \RNum{2}' before and after the normalization is compared in Fig. \ref{sic_tomo_precision} (e).

\textit{Discussion and conclusions.}--By analyzing the precision of the DQDT protocol shown in Figs. \ref{proj_rank1} (d), \ref{sic_tomo_precision} (d) and \ref{sic_tomo_precision} (e), we find that the statistical errors dramatically increases when the coupling strength $g$ approaches $0$. Therefore, one should avoid the DQDT working with a small $g$. In the characterization of rank-1 POVMs, the precision gets worse when the retrodicted state of the rank-1 POVM element approaches being orthogonal to the pre-selected state, which can be overcome with the use of completeness condition of the POVM. Though a single pre-selected state is sufficient to determine a rank-1 POVM, deviation may occur when the realistic imperfections render the POVM element with higher rank. Thus, measuring the Dirac distribution of the POVM provides the complementary verification and the improvement of the accuracy to the preceding characterization of the rank-1 POVM.

In conclusion, we propose a direct tomography scheme to characterize unknown quantum measurements by associating the POVMs with the numerators of the weak values. An appropriate choice of the coupling strength as well as the completeness condition of the POVM allows us to gain a precise characterization of the arbitrary POVMs. In the experiment, the direct tomography scheme is applied to characterize both the projective measurements and the general POVM. The DQDT results coincide well with the theoretical predictions and the results of CT, while showing the algorithmic and operational simplification in characterizing complicated measurement apparatus. Our scheme Extends the direct tomography from quantum states, quantum processes to quantum measurements, which not only provides new tools for investigating non-classical features of quantum measurement but also highlights the time-symmetric formulation of the weak measurement and extends its scope.

\begin{acknowledgments}
This work was supported by the National Key Research and Development Program of China (Grant Nos. 2017YFA0303703 and 2018YFA030602) and the National Natural Science Foundation of China (Grant Nos. 91836303, 61975077, 61490711 and 11690032) and Fundamental Research Funds for the Central
Universities (Grant No. 020214380068).
\end{acknowledgments}


%

\end{document}